\documentclass[%
 aip,
 jmp,%
 amsmath,amssymb,
 reprint,%
 floatfix,
]{revtex4-1}

\usepackage{graphicx}
\usepackage{dcolumn}
\usepackage{bm}

\usepackage{hyperref}
\usepackage{mathrsfs}

\newcommand{\Jac}{\operatorname{Jac}}
\newcommand{\Covp}{\operatorname{Cov}_p}

\newcommand{\dE}{{\,dE}}

\begin{document}


\title{Optimum Experimental Design for EGDM~Modeled~Organic~Semiconductor~Devices}

\author{C. K. F. Weiler}
\email{christoph.weiler@iwr.uni-heidelberg.de}
\affiliation{Interdisciplinary Center for Scientific Computing (IWR), Heidelberg University, 
   69120 Heidelberg, Germany.}
\author{S. K\"orkel}%
\affiliation{Interdisciplinary Center for Scientific Computing (IWR), Heidelberg University, 
   69120 Heidelberg, Germany.}

\date{\today}

\begin{abstract}
We apply optimum experimental design (OED) to organic semiconductors
modeled by the extended Gaussian disorder model (EGDM) which was developed by Pasveer et al.
\cite{Pasveer2005}
We present an extended Gummel method to decouple the corresponding system of equations and use automatic 
differentiation to get derivatives with the required accuracy for OED. We show in two
examples, whose parameters are taken from Pasveer et al.\ \cite{Pasveer2005} and 
Mensfoort and Coehoorn \cite{mensfoort-coehoorn}
that the linearized confidence regions of the parameters can be reduced significantly 
by applying OED resulting in new experiments with a different setup.
\end{abstract}

\pacs{81.05.Fb,02.60.Lj,02.60.Cb,02.60.Jh,02.60.Pn}
\keywords{Organic Semiconductor, Gummel Method, Automatic Differentiation, Optimum Experimental Design,
    Confidence Region, Covariance Matrix}
\maketitle

\section{\label{sec:1}Introduction}

Simulation of organic semiconductor devices, e.g.\ organic light emitting 
diodes (OLED), organic solar cells, etc., has gained great interest in the past decade.
Accurate models often lack in precise parameters and their identification
is time-consuming or expensive. 
One major problem is the uncertainty in the measurement data which leads 
to uncertain parameters. Carrying out more experiments would minimize this uncertainty in
an expensive way. An alternative is to use
OED in order to minimize the parameter uncertainty by planning new
(optimal) experiments. New measurement data are received
for which the parameter estimation yields parameters with minimal confidence intervals.
We apply the concept of OED to the EGDM, a special model
for the mobility of electron transport in organic polymeric material.
The chapters are arranged as follows:
We give a brief overview of the model equations in Sec. \ref{sec:2}
and point out what the relevant quantities are. In Sec. \ref{sec:3}, we explain the 
methodology of the optimum experimental design problem in more detail. After describing the
equation solver methods we used, Sec. \ref{sec:4}, numerical results of the optimization
are presented in Sec. \ref{sec:5}.

\section{\label{sec:2}EGDM equations}
  A basic description of charge transport in semiconducting materials in the steady-state case 
  is given by the coupled van Roosbroeck system\cite{Roosbroeck1950} consisting of the continuity equation, 
  also called drift-diffusion equation, and the Poisson equation.
  Given a domain $\Omega := (0,L) \subset \mathbb R$,
  the state variables, i.e.\ the space dependent functions,
  are the electric charge density $n$ in $[m^{-3}]$ and the electric potential $\phi$
  in $[eV]$ which are scalar real valued functions defined on $\Omega$.
  We assume that $n$ and $\phi$ are twice differentiable on $\Omega$.
  Pasveer et al.\ \cite{Pasveer2005} proposed the EGDM to conjugated semi-conducting polymers,
  where the diffusion and the mobility depend on the state variables 
  and hence on the space variable $x$. 
  Furthermore they introduced another state $E_F$, 
  called Quasi-Fermi-Energy, and a corresponding equation which couples $E_F$ and $n$
  at every space point.
  We omit the $x$-dependence in the following equations, only $n$, $\phi$ and $E_F$ are space
  dependent.
  The model equations are:
    \begin{align}
      0 & = \partial_x\,J(n,\phi,E_F),  \label{eq:cont-eq}\\
       -\partial_x^2 \phi & = \frac e\varepsilon \, n,   \label{eq:poiss-eq}\\
        n & =  \frac{N_t}{\sqrt{2\pi{\sigma}^2}} \int \limits_{-\infty}^\infty
           e^{-\frac{E^2}{2{\sigma}^2}}
           \frac{1}{1+\exp\left(\frac{E-E_F}{k_BT}\right)}\dE, \label{eq:fermi-eq}
    \end{align}
  where
  {\small \begin{align}
      J(n,\phi,E_F) & = {\mu_0}\,g_0\,
             g_1(n)g_2(\phi) \\
             & \qquad \Big(n\partial_x \phi -{k_B {T}             
             g_3(n,E_F)}\,\partial_x n \Big),  \label{eq:current}\\
             g_0 & = \exp\left\{-0.42 \left(\frac{\sigma}{k_B T}\right)^2\right\},
             \label{eq:g0}\\
       g_1(n) & = \exp\left[\frac12 (\hat {\sigma}^2-\hat{{\sigma}})
       \left(\min\left\{\frac{2n}{{N_t}},0.2\right\}\right)^{\delta({\sigma})}
       \right], \\
       g_2(\phi) & = \exp\left\{ 0.44(\hat{{\sigma}}^{\frac32}-2.2)\right\} \\
       & \qquad  \left[\sqrt{1+0.8\left( \min\left\{\frac{\partial_x
       \phi}{{N_t}^{\frac13}{\sigma}},2\right\}\right)^2}
               -1\right], \\
       g_3(n, E_F) & = \frac{n}{k_B {T} \frac{dn}{dE_F}}. \label{eq:g3}
  \end{align}}In these equations $J$ is the electric current density in $[\frac{A}{m^2}]$,
  $k_B$ is the Boltzmann constant in $[\frac{eV}{K}]$, $\varepsilon$ the permittivity in
  $[\frac{As}{Vm}]$,
  $e$ the elementary charge in $[As]$,
  $\hat \sigma := \frac{\sigma}{k_B T}$ and
  $\delta(\sigma) := 2\frac{\log(\hat\sigma-\hat\sigma^2)-\log\log 4}{\hat\sigma^2}$.
  In the anorganic case, the $g_i$-factors, $i=1,2,3$, would be constant. Their
  organic model is yield by comparison with the solution of the master equation. \cite{Pasveer2005}
  On the boundary $\partial \Omega = \{0,L\}$ the following conditions are imposed:
  {\small \begin{equation*}\label{eq:boundary}
    \begin{aligned}       
       n(0) & =  \frac{{N_t}}{\sqrt{2\pi{\sigma}^2}} 
                 \int \limits_{-\infty}^\infty e^{-\frac{E^2}{2\sigma}^2}  \\
       & \  
       \left[1+\exp\left(\frac{E+{\varphi_1-\sqrt{\max\{-\frac{e}{4\pi\varepsilon}\partial_x\phi(0),0\}}}}
       {k_B{T}}\right)\right]^{-1}\dE, \\
       n(L) & =  \frac{{N_t}}{\sqrt{2\pi{\sigma}^2}} 
                 \int \limits_{-\infty}^\infty
       e^{-\frac{E^2}{2{\sigma}^2}} 
        \frac{1}{1+\exp\left(\frac{E+{\varphi_2}}
         {k_B{T}}\right)}\dE, \\
       \phi(0) & = 0, \\
       \phi({L}) & = eV-\left( \varphi_2 - \varphi_1 \right),
    \end{aligned}
  \end{equation*}}where $V$ is the voltage in $[V]$ 
  and $\varphi_{1,2}$ are given energy barriers in $[eV]$.
  On the entrance, $x=0$, the energy barrier $\varphi_1$ is lowered according
  to the theory of Emtage and O'Dwyer\cite{Emtage1966} and Scott and Malliaras. \cite{Scott1999}
  For our computations we take the dimensionless form of the equations proposed by Bonham
  and Jarvis. \cite{bonham}
  For the later use we define 
  \begin{equation} \label{eq:paracontr}
    \begin{aligned}
      p &:= (\mu_0, \sigma, N_t)\in \mathbb R^3, \\
      q &:= (L, T)\in \mathbb R^2, \\
    \end{aligned}
  \end{equation}
  where $p$ are parameters of unknown numerical value given by nature.
  They have to be identified by comparing a model response to experimental data.
  $\mu_0$ is the zero temperature mobility in $[\frac{m^2}{Vs}]$, $\sigma$ is
  the width of the Gaussian distribution of the density of states in $[eV]$ and
  $N_t$ is the site density in $[m^{-3}]$.
  We assemble quantities which are adjustable by an experimenter, 
  the device length $L$ in $[m]$ and the temperature $T$ in $[K]$, in 
  the vector $q$.\cite{Badinski2011}

\section{Solution methods for the EGDM equations}
\label{sec:3}
  There are two established ways for solving the system 
  \eqref{eq:cont-eq}-\eqref{eq:fermi-eq}:\cite{knappruhstaller}
  \begin{enumerate}
    \item Apply Newton's method to the fully coupled system of equations
    \item Use Gummel's method, \cite{gummel} i.e.\ a fixed point iteration which decouples the three EGDM equations
  \end{enumerate}
  With Newton's method, one can achieve quadratic convergence.
  However, finding good starting values is not simple. 
  In the work of Knapp et al., \cite{knappruhstaller} a strategy 
  motivated by physical considerations is described.
  Another aspect is that the Jacobian, i.e.\ derivatives of the functions
  w.r.t.\ the states, is required. 
  To compute the Jacobian, the main two options are to use difference formulas or compute
  the exact derivatives. The latter can be done by hand or by using automatic differentiation (AD).
  Either way, the additional effort for computing the derivative in $n$ directions
  is at least $2n$ times the effort of each function evaluation in
  every Newton step. 
  Another possibility is to solve \eqref{eq:cont-eq}-\eqref{eq:fermi-eq}
  with Gummel's method. 
  TABLE~\ref{tab:gummel} shows the modified
  algorithm consisting of the classical system of equations \eqref{eq:cont-eq} and 
  \eqref{eq:poiss-eq} and the Quasi-Fermi energy $E_F$ defining equation \eqref{eq:fermi-eq}.
  \cite{Stodtmann2012}
  We solve the equations sequentially and insert the interim results into the next equations.  
  \begin{table}
    \caption{\label{tab:gummel}Algorithm for the fixed point iteration of Gummel 
    expanded by the EGDM-Quasi-Fermi-Equation and a derivative-free linearization.}
    \begin{ruledtabular}
      \begin{tabular}{l}
        Let $u^0:=(n^0,\phi^0,E_F^0)$ be given and choose $\|\Delta u^0\| \gg \text{TOL}$ \\
        with given error tolerance $\text{TOL}$. \\
        Set $i=0$. \\
        \textbf{while} $\big\|\Delta u^i\big\|_2 > TOL$ \\
        \hspace{0.5cm} Solve \\
        \hspace{1cm} $n^{i} = \frac{{N_t}}{\sqrt{2\pi{\sigma}^2}} 
                    \int \limits_{-\infty}^\infty e^{-\frac{E^2}{2{\sigma}^2}}
                    \frac{1}{1+\exp\left(\frac{E-E_F^{i+1}}{k_B{T}}\right)}\dE$ \\
        \hspace{0.5cm} for $E_F^{i+1}$ with Newton's method started with $E_F^i$. \\
        \hspace{0.5cm} Solve \\
        \hspace{2.5cm}  $-\partial_x^2 \phi^{i+1} = \lambda n^i$ \\
        \hspace{0.5cm} with corresponding boundary conditions for $\phi^{i+1}$. \\
        \hspace{0.5cm} With $n^i$, $\phi^{i+1}$ and $E_F^{i+1}$ solve \\
        \hspace{1cm} $0 = \partial_x \bigg\{\,g_1(n^i)g_2(\phi^{i+1}) $ \\
        \hspace{2.2cm} $\Big( n^{i+1}\,\partial_x \phi^{i+1} - 
        k_B T g_3(n^i, E_F^{i+1})\,\partial_x n^{i+1} 
                       \Big)\bigg\}$ \\
        \hspace{0.5cm} with corresponding boundary conditions for $n^{i+1}$. \\
        \hspace{0.5cm} Set $u^{i+1}:=(n^{i+1},\phi^{i+1},E_F^{i+1})$
                       and $\Delta u^{i+1}:= u^{i+1}-u^{i}$. \\
        \hspace{0.5cm} $i \leftarrow i+1$
      \end{tabular}
    \end{ruledtabular}
  \end{table}
  With the special linearization, used in the third equation of TABLE \ref{tab:gummel}, 
  we do not need any derivatives and the effort of function evaluations is limited. 
  We discretize the infinite dimensional problem \eqref{eq:cont-eq}-\eqref{eq:fermi-eq}
  to a finite one.
  The domain $\overline \Omega$ is divided in $N$ subintervals $I_i :=[x_i,x_i+h],\ x_i\in
  \{x_0=0,\dotsc,x_N=L\}$ with a constant mesh size
  $h:=\frac{L}{N}$. Finite differences are applied to the spatial derivatives, i.e.\ with respect to $x$.
  The so-called Scharfetter-Gummel scheme\cite{scharfetter-gummel} forces the function $J$, 
  defined in Eq.\eqref{eq:current}, to be constant
  on each interval $I_i$, denoted by $J_{i+\frac12}$, and provides
  an upwind stabilization, so that computation on coarse meshes is possible.
  On the interval $I_i$, the scheme looks like 
  \begin{equation*}
   J_{i+\frac12} = \mu_0\,g_0\,
      \tilde g_1 \tilde g_2
      \frac{\phi_{i+1}-\phi_i}{h}\,
      \frac{
        n_{i+1}\exp
        \left(
          -\frac{\phi_{i+1}-\phi_i}{k_B T \tilde g_3}
        \right)-n_i}{\exp
        \left(
          -\frac{\phi_{i+1}-\phi_i}{k_B T \tilde g_3}
        \right)-1}.
  \end{equation*}
  The terms $\tilde g_j$, $j=1,2,3$ stand for average values of the non-constant 
  functions $g_j$, $j=1,2,3$. It is important that the averages are taken
  in an upwind conform way to prevent numerical oscillations. 

\section{Optimum Experimental Design for model validation}
\label{sec:4}
  In this part, we follow the approaches of Lohmann \cite{Lohmann1992a} and K\"orkel et al. 
  \cite{Bauer2000} 
  With different choices of the controls $q$ and the voltage $V$, defined in Sec. \ref{sec:2}, 
  we set up multiple experiments in which the current density $J$ \eqref{eq:current} is measured.
  Let $M$ be the number of measurements we yield.
  In a parameter estimation, the parameters are identified 
  by fitting a model response, here $J$, to experimental data, i.e.\ measurements.
  We assume the measurement error to be normally distributed with mean zero and 
  covariance matrix $\Sigma^2 = \operatorname{diag}(\sigma_i^2, i=1,\dotsc, M) \in \mathbb
  R^{M\times M}$.
  With the same experimental settings, i.e.\ equal controls $q$,
  a fit from a different realization of the measurement errors may result
  in very different parameter values.
  The covariance matrix of the parameters allows to analyze the quality
  of a parameter estimation. 
  The assumed model for the standard deviations of the measurement errors is: \cite{Badinski2011}
  \begin{equation*}
    \sigma_i = 0.1 \cdot \mathcal J_i + 0.1 \quad \left[\frac{A}{m^2}\right],
  \end{equation*}
  where $\mathcal J_i \in\mathbb R$ is the function value of $J$ corresponding to the $i$-th measurement.
  For further notation, we assemble the values $\mathcal J_i$ in the vector $\mathcal J \in \mathbb
  R^M$.
  If the confidence region of the parameters is approximated by assuming a linear propagation of the 
  measurement errors, 
  it can be parameterized by the covariance matrix defined by
  \begin{equation*}
    \Covp := \mathbb E[(p-\mathbb E[p])(p-\mathbb E[p])^T] \in \mathbb R^{N_p\times N_p},
  \end{equation*}
  where $N_p$ is the number of parameters.
  From now on, we denote by $n$, $\phi$ and $E_F$ the discrete counterparts of the 
  state variables which are  $(N-1)$-dimensional vectors, without boundary values. 
  They assemble the function values at the mesh points $x_i$, $i=1,\dotsc,N-1$, cf.\ Sec.\ref{sec:3}.
  We abbreviate the discrete solution of the system \eqref{eq:cont-eq}-\eqref{eq:fermi-eq}
  dependent on parameters $p$ and controls $q$ by 
  \begin{equation*}
     u(p,q) := (n(p,q),\phi(p,q),E_{F}(p,q)) \in \mathbb R^{N_u},
  \end{equation*}
  cf.\ TABLE \ref{tab:gummel}. We used the notation $N_u:=3(N-1)$ for the overall
  state dimension. In the following we denote the derivative of a function 
  $f$ w.r.t.\
  $x$ in the direction $\Delta x$ by 
  \begin{equation*}
    d_x f \{\Delta x\} := \lim_{\varepsilon\to 0} \frac{f(x+\varepsilon \Delta x)}{\varepsilon}
  \end{equation*}
  and write accordingly the second derivatives of $f$ 
  w.r.t.\ $x$ and $y$ in the directions
  $\Delta x$ and $\Delta y$ as $d^2_{yx} f \{\Delta x,\Delta y\}$.
  We combine several directions $\Delta x_i$ into a so-called seed matrix 
  $S = (\Delta x_1, \cdots, \Delta x_n)$ and define $d_x f \{S\}$ by
  \begin{equation*}
    \left(d_x f \{S\}\right)_{ij} = d_x f_i \{\Delta x_j\}
  \end{equation*}
  and accordingly $d^2_{yx} f \{S_1,S_2\}$ by
  \begin{equation*}
    (d^2_{yx} f \{S_1,S_2\})_{ijk} = d^2_{yx} f_i \{\Delta x_j,\Delta y_k\}
  \end{equation*}
  with $S_1 =  (\Delta x_j)_j$ and $S_2 =  (\Delta y_k)_k$.
  We define the $M\times N_p$ Matrix
  \begin{equation*}
    \begin{aligned}
      \Jac(u(p,q),p,q) & := -\Sigma^{-1}d_p \mathcal J(u(p,q),p,q) \{I_p\} \\
      & \ =  -\Sigma^{-1} \bigg[ \partial_u \mathcal J \{\partial_p u \{I_p\}\}
      + \partial_p \mathcal J \{ I_p\}\bigg],
    \end{aligned}
  \end{equation*}
  with the derivative of $\mathcal J$ w.r.t.\ to the parameters $p$ with the 
  $N_p$-dimensional identity matrix $I_p$ as seed matrix.
  Computing 
  $\partial_u \mathcal J \{\partial_p u \{I_p\}\}$ is much less expensive than
  computing the matrix product
  $\partial_u \mathcal J\{I_u\}\cdot \partial_p u \{I_p\}$
  with the identity matrix $I_u\in\mathbb R^{N_u \times N_u}$. 
  The covariance matrix in the unconstrained case can be computed by
  \begin{equation}\label{eq:covmat}
    \Covp = \big(\Jac(u(p,q),p,q)^T \Jac(u(p,q),p,q)\big)^{-1}.
  \end{equation}
  For given probability $\alpha \in [0,1]$ the linearized $(100\cdot \alpha)\%$-confidence region is 
  described by
  \begin{equation}\label{eq:conf_reg_eq}
    G(\alpha,p) = \{v \in \mathbb R^{N_p} : v = p + \delta p, \delta p^T \Covp^{-1} \delta p \leq
    \gamma(\alpha)^2\},
  \end{equation}
  with the $(1-\alpha)$-quantile of the $\chi^2$-distribution $\gamma(\alpha)^2$.
  As an approximation of the confidence intervals of the parameters we use
  \begin{equation}\label{eq:conf-interval}
    \begin{aligned}
      & [p_i - \theta_i,p_i + \theta_i] \quad \text{with} \\
      & \theta_i = \gamma(\alpha)\sqrt{(\Covp)_{ii}}, \quad i=1,\dotsc,N_p.
    \end{aligned}
  \end{equation}
  We point out that for 
  computing $\Covp$ (and the confidence intervals) no measurement data is necessary.
  We consider the reduced nonlinear optimization problem
  \begin{equation}\label{eq:oedproblem}
    \begin{aligned}
      \min_{q} & \quad 
      \frac{1}{N_p}\operatorname{trace}\big(\Covp\big)  \\
      \end{aligned}
  \end{equation}
  which is called ``optimum experimental design problem''.  
  Efficient optimization algorithms require derivative information.
  For higher order difference formulas, the number of function evaluations 
  increases which can be very costly like in our case.
  Additionally, low order schemes reduce the number of exact digits, 
  in particular in combination with derivatives of higher order.
  To avoid such approximation errors, we use automatic differentiation
  where the derivatives are evaluated up to machine precision.
  For derivative based optimization, we need the derivative of the gradient
  of $\Jac$ with the seed matrix $I_q\in\mathbb R^{N_q \times N_q}$, 
  where $N_q$ is the number of the controls.
  $N_p$ and $N_q$ are relatively small, so we use the forward mode, where we have 
  to compute
  \begin{equation*}
    \begin{aligned}
      & d_q\Jac(u(p,q),p,q) \left\{I_q\right\} \\
      & \ = -\Sigma^{-1}  \Big[ d_q\big( \partial_u \mathcal J \{\partial_p u \{I_p\}\}
         + \partial_p \mathcal J \{I_p\} \big) \{I_q\} \Big] \\
         & \ = -\Sigma^{-1} \Big[ \partial^2_{u}\mathcal J\{\partial_p u \{I_p\},
         \partial_q u \{I_q\}\} + \partial^2_{qu}\mathcal J\{\partial_p u \{I_p\},I_q\} \\
         & \ \qquad\qquad  + \partial_u \mathcal J \big\{\partial^2_{qp}u \{I_p,I_q\} \big\} 
         + \partial^2_{up}\mathcal J\{I_p,\partial_q u \{I_q\}\} \\
         & \ \qquad\qquad + \partial^2_{qp}\mathcal J\{I_p,I_q\} \Big],
    \end{aligned}
  \end{equation*}
  see e.g.\ Griewank. \cite{Griewank2000}
  We need derivatives of $u$ w.r.t.\ the parameters $p$, controls
  $q$ and mixed second derivatives w.r.t.\ $p$ and $q$.  
  One way would be to differentiate the fixed point iteration, TABLE \ref{tab:gummel}, and 
  iterate over the derivatives as well. \cite{Christianson1994}
  However, we are only interested in the derivatives of the solution of 
  Gummel's method $u(p,q)$. In the solution it holds
  \begin{equation}\label{eq:ststsh}
    F(u(p,q),p,q) = 0  \qquad \forall p,q,
  \end{equation}
  with $F$ the discrete counterpart of the 
  system of equations \eqref{eq:cont-eq}-\eqref{eq:fermi-eq}.
  To compute the required derivatives $\partial_p u\{I_p\}$, $\partial_q u\{I_q\}$
  and $\partial^2_{qp} u\{I_p,I_q\}$, we use the sensitivity method, 
  see e.g.\ Hinze et al. \cite{Hinze2010}
  Using adjoint method is not recommended here, because
  the number of measurements $M$ is much higher than the number of parameters $N_p$.
  Differentiating \eqref{eq:ststsh} leads to
  \begin{equation*}
    \begin{aligned}
      d_p F(u(p,q),p,q)\{I_p\} = 0, \\
      d_q F(u(p,q),p,q)\{I_q\} = 0, \\
      d^2_{qp} F(u(p,q),p,q)\{I_p,I_q\} = 0. \\
    \end{aligned}
  \end{equation*}
  Hence the required derivatives are given in form of
  \begin{equation}\label{eq:sensitivities}
      \begin{aligned}
        {\partial_p u}\{I_p\} & = -\big(\partial_u F\{I_u\}\big)^{-1}\partial_p F\{I_p\} \\
        {\partial_q u}\{I_q\} & = -\big(\partial_u F\{I_u\}\big)^{-1}\partial_q F\{I_q\} \\
        {\partial^2_{qp} u} \{I_p,I_q\} & = 
             -\left(\partial_u F\{I_u\}\right)^{-1} \Big(
             \partial^2_{qp}F\{I_p,I_q\} \\
             & \hspace{-0.85cm}   + \partial^2_{up}F\left\{I_p,\partial_qu \{I_q\}\right\} + 
             \partial^2_{qu}F\left\{ \partial_pu \{I_p\}, I_q\right\} \\
             &  \hspace{-0.85cm} + \partial_u^2 F \left\{\partial_pu \{I_p\},\partial_pu \{I_q\}\right\} \Big).
    \end{aligned}
  \end{equation}
  We can save the decomposition of the matrix $\partial_u F\{I_u\}$ and
  use it for all three equations to reduce complexity.

\section{Numerical results}
\label{sec:5} 
\begin{figure}[htbp]
    \centering
      \includegraphics[width=\linewidth]{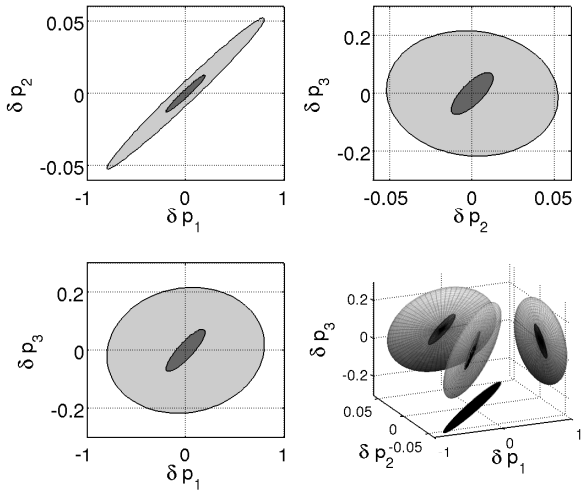}
      \caption{Projections and three dimensional ellipsoid with shadows of the linearized $95\%$-confidence 
               regions before (light part) and after (dark part) the optimization. Computed with
               Pasveer parameters \eqref{eq:pasveer_paras}.}
      \label{fig:1}
  \end{figure}

\begin{figure}[htbp]
      \includegraphics[angle=270,width=\linewidth]{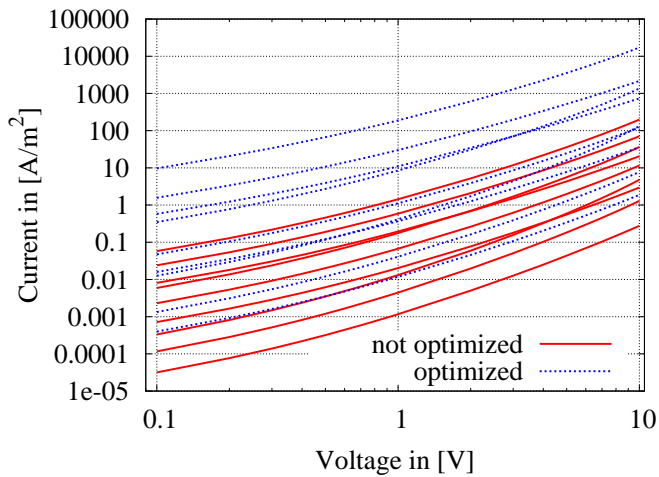}
      \caption{Comparison of current-voltage characteristics before and after the
      optimization for Pasveer parameters \eqref{eq:pasveer_paras}. 
      The continuous lines correspond to experiments with configuration $\mathcal L$
      and $\mathcal T$ and the dashed lines correspond to experiments with optimal 
      configuration $\mathcal L^\ast$ and $\mathcal T^\ast$.}
      \label{fig:2}                                              
  \end{figure}

\begin{figure}[htbp]
      \centering
      \includegraphics[width=\linewidth]{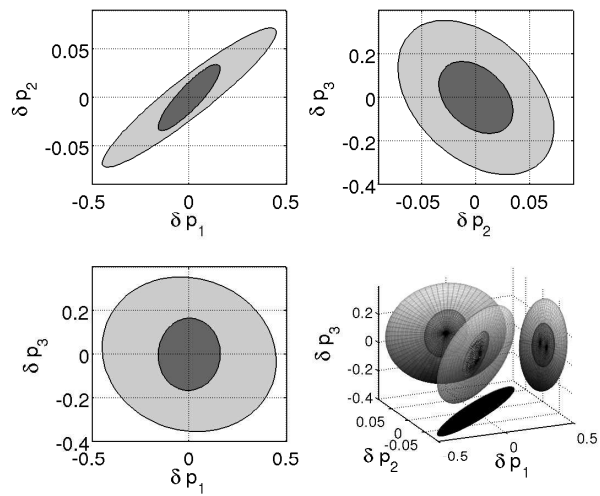}
      \caption{Projections and three dimensional ellipsoid with shadows of the linearized $95\%$-confidence 
               regions before (light part) and after (dark part) the optimization. Computed with
               Coehoorn parameters \eqref{eq:coehoorn_paras}.}
      \label{fig:3}
  \end{figure}

\begin{figure}[htbp]
      \includegraphics[angle=270,width=\linewidth]{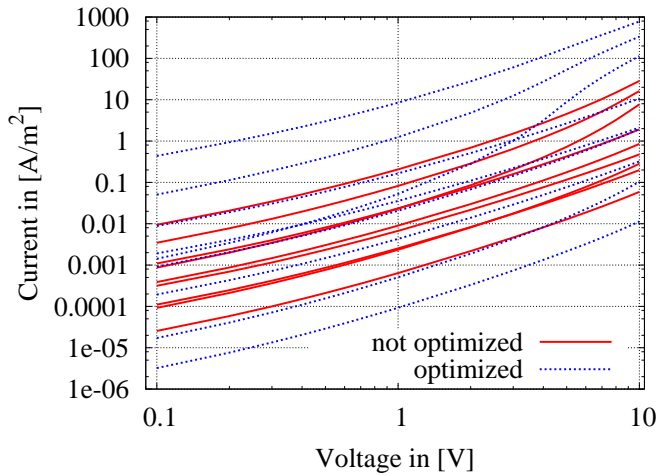}
      \caption{Comparison of current-voltage characteristics before and after the
      optimization for Pasveer parameters \eqref{eq:coehoorn_paras}. 
      The continuous lines correspond to experiments with configuration $\mathcal L$
      and $\mathcal T$ and the dashed lines correspond to experiments with optimal 
      configuration $\mathcal L^\ast$ and $\mathcal T^\ast$.}
      \label{fig:4}
  \end{figure}

  Experimenters proceed in the following way.\cite{AlHelwi2011} 
  Devices of different lengths $L$ are produced by evaporating.
  At different temperatures $T$, they apply voltage series $\mathcal V$, a vector consisting 
  of various voltages $V$, and measure the corresponding current values, combined in the vector $\mathcal J$. 
  One choice of $L$, $T$ and voltage series $\mathcal V$ we call one experiment 
  subsequently.
  We will show in two examples how the confidence regions \eqref{eq:conf_reg_eq}
  can be reduced by OED. In each example we consider three different lengths 
  \begin{equation*}
    \begin{aligned}
       \mathcal L & := \left(L_1,L_2,L_3\right)
    \end{aligned}
  \end{equation*}
  and three different temperatures 
  \begin{equation*}
    \begin{aligned}
       \mathcal T & := \left(T_1,T_2,T_3\right).
    \end{aligned}
  \end{equation*}
  The combination of each element of the vector $\mathcal L$ with each element of the vector $\mathcal T$ lead
  to nine experiments. We assemble all control variables, which are the optimization variables
  for the OED problem, in a vector 
  \begin{equation*}
    \mathcal Q := \big( L_1,T_1,L_1,T_2,L_1,T_3,L_2,T_2\dotsc,L_3,T_3\big) \in \mathbb R^{18}.
  \end{equation*}
  The result of the OED should be three lengths and three temperatures combined in
  nine experiments. To enforce that, we apply additional constraints to $\mathcal Q$
  for the lengths
  \begin{equation*}\label{eq:cfcn}
    \begin{aligned}
      \mathcal Q_1 & = \mathcal Q_3 = \mathcal Q_5, \\
      \mathcal Q_7 & = \mathcal Q_9 = \mathcal Q_{11}, \\
      \mathcal Q_{13} & = \mathcal Q_{15} = \mathcal Q_{17},
    \end{aligned}
  \end{equation*}
  and for the temperatures
  \begin{equation*}
    \begin{aligned}
      \mathcal Q_{2} & = \mathcal Q_{8} = \mathcal Q_{14}, \\
      \mathcal Q_{4} & = \mathcal Q_{10} = \mathcal Q_{16}, \\
      \mathcal Q_{6} & = \mathcal Q_{12} = \mathcal Q_{18}. \\
    \end{aligned}
  \end{equation*}
  We solve problem \eqref{eq:oedproblem} with the software package {\it VPLAN}\cite{Koerkel2002} 
  where the optimization problem is solved with an inexact SQP method 
  provided by {\it SNOPT 7.2-9 (Jun 2008)}. \cite{Gill2002}
  In each iteration, the objective, i.e.\ the covariance matrix, is computed
  by solving the underlying system of equations.
  We implemented the extended Gummel method TABLE~\ref{tab:gummel} in {\it VPLAN} to solve
  this system. For the optimization, the derivatives of the model functions are calculated with the 
  AD tool {\it ADIFOR 2.0}. \cite{Bischof1994} We implemented a routine in {\it VPLAN}, which solves
  the system of equations \eqref{eq:sensitivities}.
  In a first example we take configurations $p$, $\mathcal V$, $\mathcal L$ and 
  $\mathcal T$ from Pasveer et al.\ \cite{Pasveer2005} who used
  \begin{equation}\label{eq:pasveer_paras}
    \begin{aligned}
       p & =\left(1.15\cdot 10^{-5},0.14,2.44\cdot 10^{26} \right), \\
       \mathcal V & = (0.1,\dotsc,0.9,1,2,\dotsc,10), \\
       \mathcal L & = \left(275,350,475\right), \\
       \mathcal T & = \left(235,270,305\right).
    \end{aligned}
  \end{equation}
  Here and in the following example we set $\varphi_1 = \varphi_2 = 0$ for the energy barriers 
  in $[eV]$ and $\varepsilon = 2.66\cdot 10^{-11}$ for the permittivity in $[\frac{As}{Vm}]$.
  We choose boundaries to the lengths and temperatures according to the practicability
  of experiments and the validity of the model:
  \begin{equation}\label{eq:bounds}
    L_i\in[50,500] \quad \text{and} \quad T_i\in[200,350] \quad i=1,2,3.
  \end{equation}  
  The optimization results in the vectors
  \begin{align*}
    \mathcal L^\ast & = (50,339.1,471.6), \\
    \mathcal T^\ast & = (277.2,281.8,350).
  \end{align*}
  Where the lengths do not change much, a tendency to higher temperatures is
  observed. The algorithm gives back the exact numbers of the optimum. 
  Dependent on the equipment of the laboratory these values can not be 
  realized in practice. Nevertheless they can be taken as a guideline. 
  All the values in the neighborhood of $\mathcal L^\ast$ and $\mathcal T^\ast$
  are leading to similar results.  
  In Fig. \ref{fig:1}, we visualize the three dimensional ellipsoid and the projections corresponding
  to the set $G_L(0.95,0)$ before and after the optimization.
  TABLE~\ref{tab:2} shows a comparison of the confidence intervals
  of the parameters. The average of the squared semi-axes of the optimized ellipsoid,
  i.e.\ the objective in \eqref{eq:oedproblem}, is $0.07$ times the 
  average of the squared semi-axes of the ellipsoid not optimized.
  \begin{table}
    \caption{\label{tab:2}Radii of the confidence intervals 
    before and after the optimization computed
    by \eqref{eq:conf-interval} with Pasveer parameters \eqref{eq:pasveer_paras}.}
    \begin{tabular}{p{0.7cm}rr}
       \hline\hline \\[-0.3cm]
        & $\theta_{\text{start}}$ & $\theta_{\text{optimal}}$ \\
       \hline \\[-0.3cm]
          ${ p_1}$ & 79.90\% & 19.85\% \\ 
          ${ p_2}$ &  5.21\% &  1.27\% \\ 
          ${ p_3}$ & 21.62\% &  7.21\% \\ 
       \hline\hline
     \end{tabular}
  \end{table}
  The results of the simulations before and 
  after the optimization procedure are shown in Fig. \ref{fig:2}.
  The optimal configuration leads to higher current densities.
  A second example is taken from Mensfoort and Coehoorn, \cite{mensfoort-coehoorn}
  who used 
  \begin{equation}\label{eq:coehoorn_paras}
    \begin{aligned}
       p & =\left( 1.0\cdot 10^{-10},0.077, 4.25\cdot 10^{26} \right), \\
       \mathcal V & = \left( 0.1,\dotsc,0.9,1,2,\dotsc,10 \right), \\
       \mathcal L & = \left(100,200,300\right), \\
       \mathcal T & = \left(235,270,305\right).
    \end{aligned}
  \end{equation}
  Again we take boundaries like in \eqref{eq:bounds}
  and the resulting vectors are   
  \begin{align*}
    \mathcal L^\ast & = (50,187.8,296.8), \\
    \mathcal T^\ast & = (200,274.7,350).
  \end{align*}
  Unlike the first example, the optimal lengths differ much more from the starting values
  and not all temperatures were raised, the lowest temperature was even reduced.
  In Fig. \ref{fig:3}, we illustrate the confidence ellipsoid and the projections and 
  TABLE~\ref{tab:3} shows a comparison of the confidence intervals in this case.
  The average of the squared semi-axes of the optimized ellipsoid here is $0.16$ times the 
  average of the squared semi-axes of the ellipsoid not optimized, too.  
  \begin{table}
    \caption{\label{tab:3}Radii of the confidence intervals 
    before and after the optimization computed
    by \eqref{eq:conf-interval} with Coehoorn parameters \eqref{eq:coehoorn_paras}.}
    \begin{tabular}{p{0.7cm}rr}
       \hline\hline \\[-0.3cm]
        & $\theta_{\text{start}}$ & $\theta_{\text{optimal}}$ \\
       \hline \\[-0.3cm]
            ${p_1}$ & 44.76\%  & 15.98\% \\ 
            ${p_2}$ &  7.20\%  &  3.42\% \\ 
            ${p_3}$ & 35.42\%  & 16.51\% \\
       \hline\hline
     \end{tabular}
  \end{table}
  The results of the simulations before and 
  after the optimization procedure are shown in Fig. \ref{fig:4}.
  The optimized configuration leads to a wider range of current densities gaining
  more information.

\section{Conclusion and Outlook}
We have applied optimum experimental design to~organic 
semiconductors modeled by the EGDM in order to reduce the parameter uncertainty caused
by measurement errors. The variances of the parameters, i.e.\ the average of the squared 
semi-axes of the confidence ellipsoid, were $0.07$ and $0.16$ 
times the average of the squared semi-axes of the not optimized confidence ellipsoid 
respectively. The proposed experiments, followed by a parameter estimation 
lead to a model, which is approximately $10$ times more exact with respect to the assumed 
measurement error than the previous one. 
The classical methods for solving semiconductor models were 
extended to fit with the EGDM. The derivatives required for the optimization were
computed via automatic differentiation, leaving the error on machine precision level, 
even for higher derivatives. We used unipolar layer devices for simplicity, but
the presented methods can also be applied to multi-layer devices, trap generation models 
and exciton rate equations or, more far-reaching, all other models which are based on the 
van Roosbroeck system.
\\

\begin{acknowledgments}
  The authors acknowledge financial support from the
  German Federal Ministry of Education and Research (BMBF),
  project PARAPLUE 03MS649A.
\end{acknowledgments}


\begin{thebibliography}{20}%
\makeatletter
\providecommand \@ifxundefined [1]{%
 \@ifx{#1\undefined}
}%
\providecommand \@ifnum [1]{%
 \ifnum #1\expandafter \@firstoftwo
 \else \expandafter \@secondoftwo
 \fi
}%
\providecommand \@ifx [1]{%
 \ifx #1\expandafter \@firstoftwo
 \else \expandafter \@secondoftwo
 \fi
}%
\providecommand \natexlab [1]{#1}%
\providecommand \enquote  [1]{``#1''}%
\providecommand \bibnamefont  [1]{#1}%
\providecommand \bibfnamefont [1]{#1}%
\providecommand \citenamefont [1]{#1}%
\providecommand \href@noop [0]{\@secondoftwo}%
\providecommand \href [0]{\begingroup \@sanitize@url \@href}%
\providecommand \@href[1]{\@@startlink{#1}\@@href}%
\providecommand \@@href[1]{\endgroup#1\@@endlink}%
\providecommand \@sanitize@url [0]{\catcode `\\12\catcode `\$12\catcode
  `\&12\catcode `\#12\catcode `\^12\catcode `\_12\catcode `\%12\relax}%
\providecommand \@@startlink[1]{}%
\providecommand \@@endlink[0]{}%
\providecommand \url  [0]{\begingroup\@sanitize@url \@url }%
\providecommand \@url [1]{\endgroup\@href {#1}{\urlprefix }}%
\providecommand \urlprefix  [0]{URL }%
\providecommand \Eprint [0]{\href }%
\providecommand \doibase [0]{http://dx.doi.org/}%
\providecommand \selectlanguage [0]{\@gobble}%
\providecommand \bibinfo  [0]{\@secondoftwo}%
\providecommand \bibfield  [0]{\@secondoftwo}%
\providecommand \translation [1]{[#1]}%
\providecommand \BibitemOpen [0]{}%
\providecommand \bibitemStop [0]{}%
\providecommand \bibitemNoStop [0]{.\EOS\space}%
\providecommand \EOS [0]{\spacefactor3000\relax}%
\providecommand \BibitemShut  [1]{\csname bibitem#1\endcsname}%
\let\auto@bib@innerbib\@empty
\bibitem [{\citenamefont {Pasveer}\ \emph {et~al.}(2005)\citenamefont
  {Pasveer}, \citenamefont {Cottaar}, \citenamefont {Tanase}, \citenamefont
  {Coehoorn}, \citenamefont {P.~A.~Bobbert}, \citenamefont {de~Leeuw},\ and\
  \citenamefont {Michels}}]{Pasveer2005}%
  \BibitemOpen
  \bibfield  {author} {\bibinfo {author} {\bibfnamefont {W.~F.}\ \bibnamefont
  {Pasveer}}, \bibinfo {author} {\bibfnamefont {J.}~\bibnamefont {Cottaar}},
  \bibinfo {author} {\bibfnamefont {C.}~\bibnamefont {Tanase}}, \bibinfo
  {author} {\bibfnamefont {R.}~\bibnamefont {Coehoorn}}, \bibinfo {author}
  {\bibfnamefont {P.~W. M.~B.}\ \bibnamefont {P.~A.~Bobbert}}, \bibinfo
  {author} {\bibfnamefont {D.~M.}\ \bibnamefont {de~Leeuw}}, \ and\ \bibinfo
  {author} {\bibfnamefont {M.~A.~J.}\ \bibnamefont {Michels}},\ }\href@noop {}
  {\bibfield  {journal} {\bibinfo  {journal} {Phys. Rev. Lett.}\ }\textbf
  {\bibinfo {volume} {94}} (\bibinfo {year} {2005})}\BibitemShut {NoStop}%
\bibitem [{\citenamefont {van Mensfoort}\ and\ \citenamefont
  {Coehoorn}(2008)}]{mensfoort-coehoorn}%
  \BibitemOpen
  \bibfield  {author} {\bibinfo {author} {\bibfnamefont {S.~L.~M.}\
  \bibnamefont {van Mensfoort}}\ and\ \bibinfo {author} {\bibfnamefont
  {R.}~\bibnamefont {Coehoorn}},\ }\href@noop {} {\bibfield  {journal}
  {\bibinfo  {journal} {Phys.\ Rev.\ B}\ }\textbf {\bibinfo {volume} {78}}
  (\bibinfo {year} {2008})}\BibitemShut {NoStop}%
\bibitem [{\citenamefont {van Roosbroeck}(1950)}]{Roosbroeck1950}%
  \BibitemOpen
  \bibfield  {author} {\bibinfo {author} {\bibfnamefont {W.}~\bibnamefont {van
  Roosbroeck}},\ }\href@noop {} {\bibfield  {journal} {\bibinfo  {journal}
  {Bell\ System\ Tech}\ }\textbf {\bibinfo {volume} {29}},\ \bibinfo {pages}
  {560} (\bibinfo {year} {1950})}\BibitemShut {NoStop}%
\bibitem [{\citenamefont {Emtage}\ and\ \citenamefont
  {O'Dwyer}(1966)}]{Emtage1966}%
  \BibitemOpen
  \bibfield  {author} {\bibinfo {author} {\bibfnamefont {P.}~\bibnamefont
  {Emtage}}\ and\ \bibinfo {author} {\bibfnamefont {J.~J.}\ \bibnamefont
  {O'Dwyer}},\ }\href@noop {} {\bibfield  {journal} {\bibinfo  {journal}
  {Phys.\ Rev.\ Lett.}\ }\textbf {\bibinfo {volume} {16}},\ \bibinfo {pages}
  {356} (\bibinfo {year} {1966})}\BibitemShut {NoStop}%
\bibitem [{\citenamefont {Scott}\ and\ \citenamefont
  {Malliaras}(1999)}]{Scott1999}%
  \BibitemOpen
  \bibfield  {author} {\bibinfo {author} {\bibfnamefont {J.~C.}\ \bibnamefont
  {Scott}}\ and\ \bibinfo {author} {\bibfnamefont {G.~G.}\ \bibnamefont
  {Malliaras}},\ }\href@noop {} {\bibfield  {journal} {\bibinfo  {journal}
  {Chem.\ Phys.\ Lett.}\ }\textbf {\bibinfo {volume} {299}},\ \bibinfo {pages}
  {115} (\bibinfo {year} {1999})}\BibitemShut {NoStop}%
\bibitem [{\citenamefont {Bonham}\ and\ \citenamefont {Jarvis}(1977)}]{bonham}%
  \BibitemOpen
  \bibfield  {author} {\bibinfo {author} {\bibfnamefont {J.}~\bibnamefont
  {Bonham}}\ and\ \bibinfo {author} {\bibfnamefont {D.}~\bibnamefont
  {Jarvis}},\ }\href@noop {} {\bibfield  {journal} {\bibinfo  {journal} {Aust.
  J. Chem.}\ }\textbf {\bibinfo {volume} {30}},\ \bibinfo {pages} {705}
  (\bibinfo {year} {1977})}\BibitemShut {NoStop}%
\bibitem [{\citenamefont {Badinski}(2011)}]{Badinski2011}%
  \BibitemOpen
  \bibfield  {author} {\bibinfo {author} {\bibfnamefont {A.}~\bibnamefont
  {Badinski}},\ }\href@noop {} {} (\bibinfo {year} {2011}),\ \bibinfo {note}
  {private communication, {BASF SE} {L}udwigshafen}\BibitemShut {NoStop}%
\bibitem [{\citenamefont {Knapp}\ \emph {et~al.}(2010)\citenamefont {Knapp},
  \citenamefont {H{\"{a}}usermann}, \citenamefont {Schwarzenbach},\ and\
  \citenamefont {Ruhstaller}}]{knappruhstaller}%
  \BibitemOpen
  \bibfield  {author} {\bibinfo {author} {\bibfnamefont {E.}~\bibnamefont
  {Knapp}}, \bibinfo {author} {\bibfnamefont {R.}~\bibnamefont
  {H{\"{a}}usermann}}, \bibinfo {author} {\bibfnamefont {H.~U.}\ \bibnamefont
  {Schwarzenbach}}, \ and\ \bibinfo {author} {\bibfnamefont {B.}~\bibnamefont
  {Ruhstaller}},\ }\href@noop {} {\bibfield  {journal} {\bibinfo  {journal}
  {Journal of applied physics}\ }\textbf {\bibinfo {volume} {108}} (\bibinfo
  {year} {2010})}\BibitemShut {NoStop}%
\bibitem [{\citenamefont {Gummel}(1964)}]{gummel}%
  \BibitemOpen
  \bibfield  {author} {\bibinfo {author} {\bibfnamefont {H.~K.}\ \bibnamefont
  {Gummel}},\ }\href@noop {} {\bibfield  {journal} {\bibinfo  {journal} {IEEE
  Trans. Electron Devices}\ }\textbf {\bibinfo {volume} {11}},\ \bibinfo
  {pages} {455} (\bibinfo {year} {1964})}\BibitemShut {NoStop}%
\bibitem [{\citenamefont {Stodtmann}\ \emph {et~al.}(2012)\citenamefont
  {Stodtmann}, \citenamefont {Lee}, \citenamefont {Weiler},\ and\ \citenamefont
  {Badinski}}]{Stodtmann2012}%
  \BibitemOpen
  \bibfield  {author} {\bibinfo {author} {\bibfnamefont {S.}~\bibnamefont
  {Stodtmann}}, \bibinfo {author} {\bibfnamefont {R.~M.}\ \bibnamefont {Lee}},
  \bibinfo {author} {\bibfnamefont {C.~K.~F.}\ \bibnamefont {Weiler}}, \ and\
  \bibinfo {author} {\bibfnamefont {A.}~\bibnamefont {Badinski}},\ }\bibfield
  {title} {\enquote {\bibinfo {title} {Numerical simulation of organic
  semiconductor devices with high carrier densities},}\ }\href@noop {}
  {\bibfield  {journal} {\bibinfo  {journal} {Journal of applied physics}\ }
  (\bibinfo {year} {2012})},\ \bibinfo {note} {submited},\ \Eprint
  {http://arxiv.org/abs/1208.3365} {arXiv:1208.3365 [physics.comp-ph]}
  \BibitemShut {NoStop}%
\bibitem [{\citenamefont {Scharfetter}\ and\ \citenamefont
  {Gummel}(1969)}]{scharfetter-gummel}%
  \BibitemOpen
  \bibfield  {author} {\bibinfo {author} {\bibfnamefont {D.~L.}\ \bibnamefont
  {Scharfetter}}\ and\ \bibinfo {author} {\bibfnamefont {H.~K.}\ \bibnamefont
  {Gummel}},\ }\href@noop {} {\bibfield  {journal} {\bibinfo  {journal} {IEEE
  Trans. Electron Devices}\ }\textbf {\bibinfo {volume} {16}},\ \bibinfo
  {pages} {64} (\bibinfo {year} {1969})}\BibitemShut {NoStop}%
\bibitem [{\citenamefont {Lohmann}, \citenamefont {Bock},\ and\ \citenamefont
  {Schl\"oder}(1992)}]{Lohmann1992a}%
  \BibitemOpen
  \bibfield  {author} {\bibinfo {author} {\bibfnamefont {T.}~\bibnamefont
  {Lohmann}}, \bibinfo {author} {\bibfnamefont {H.}~\bibnamefont {Bock}}, \
  and\ \bibinfo {author} {\bibfnamefont {J.}~\bibnamefont {Schl\"oder}},\
  }\bibfield  {title} {\enquote {\bibinfo {title} {Numerical methods for
  parameter estimation and optimal experimental design in chemical reaction
  systems},}\ }\href@noop {} {\bibfield  {journal} {\bibinfo  {journal}
  {Industrial and Engineering Chemistry Research}\ }\textbf {\bibinfo {volume}
  {31}},\ \bibinfo {pages} {54--57} (\bibinfo {year} {1992})}\BibitemShut
  {NoStop}%
\bibitem [{\citenamefont {Bauer}\ \emph {et~al.}(2000)\citenamefont {Bauer},
  \citenamefont {Bock}, \citenamefont {K\"orkel},\ and\ \citenamefont
  {Schl\"oder}}]{Bauer2000}%
  \BibitemOpen
  \bibfield  {author} {\bibinfo {author} {\bibfnamefont {I.}~\bibnamefont
  {Bauer}}, \bibinfo {author} {\bibfnamefont {H.}~\bibnamefont {Bock}},
  \bibinfo {author} {\bibfnamefont {S.}~\bibnamefont {K\"orkel}}, \ and\
  \bibinfo {author} {\bibfnamefont {J.}~\bibnamefont {Schl\"oder}},\ }\bibfield
   {title} {\enquote {\bibinfo {title} {Numerical methods for optimum
  experimental design in {DAE} systems},}\ }\href@noop {} {\bibfield  {journal}
  {\bibinfo  {journal} {J. Comput. Appl. Math.}\ }\textbf {\bibinfo {volume}
  {120}},\ \bibinfo {pages} {1--15} (\bibinfo {year} {2000})}\BibitemShut
  {NoStop}%
\bibitem [{\citenamefont {Griewank}(2000)}]{Griewank2000}%
  \BibitemOpen
  \bibfield  {author} {\bibinfo {author} {\bibfnamefont {A.}~\bibnamefont
  {Griewank}},\ }\href@noop {} {\emph {\bibinfo {title} {Evaluating
  Derivatives, Principles and Techniques of Algorithmic Differentiation}}},\
  \bibinfo {series} {Frontiers in Applied Mathematics}\ No.~\bibinfo {number}
  {19}\ (\bibinfo  {publisher} {SIAM},\ \bibinfo {address} {Philadelphia},\
  \bibinfo {year} {2000})\BibitemShut {NoStop}%
\bibitem [{\citenamefont {Christianson}(1994)}]{Christianson1994}%
  \BibitemOpen
  \bibfield  {author} {\bibinfo {author} {\bibfnamefont {B.}~\bibnamefont
  {Christianson}},\ }\bibfield  {title} {\enquote {\bibinfo {title} {Reverse
  accumulation and attractive fixed points},}\ }\href@noop {} {\bibfield
  {journal} {\bibinfo  {journal} {Optim. Methods Soft.}\ }\textbf {\bibinfo
  {volume} {3}},\ \bibinfo {pages} {311--326} (\bibinfo {year}
  {1994})}\BibitemShut {NoStop}%
\bibitem [{\citenamefont {Hinze}\ \emph {et~al.}(2010)\citenamefont {Hinze},
  \citenamefont {Pinnau}, \citenamefont {Ulbrich},\ and\ \citenamefont
  {Ulbrich}}]{Hinze2010}%
  \BibitemOpen
  \bibfield  {author} {\bibinfo {author} {\bibfnamefont {M.}~\bibnamefont
  {Hinze}}, \bibinfo {author} {\bibfnamefont {R.}~\bibnamefont {Pinnau}},
  \bibinfo {author} {\bibfnamefont {M.}~\bibnamefont {Ulbrich}}, \ and\
  \bibinfo {author} {\bibfnamefont {S.}~\bibnamefont {Ulbrich}},\ }\href
  {http://books.google.de/books?id=GYPAYgEACAAJ} {\emph {\bibinfo {title}
  {Optimization with PDE Constraints}}},\ Mathematical Modelling: Theory and
  Applications\ (\bibinfo  {publisher} {Springer},\ \bibinfo {year}
  {2010})\BibitemShut {NoStop}%
\bibitem [{\citenamefont {Al-Helwi}(2011)}]{AlHelwi2011}%
  \BibitemOpen
  \bibfield  {author} {\bibinfo {author} {\bibfnamefont {M.}~\bibnamefont
  {Al-Helwi}},\ }\href@noop {} {} (\bibinfo {year} {2011}),\ \bibinfo {note}
  {private communication, {BASF SE} {L}udwigshafen}\BibitemShut {NoStop}%
\bibitem [{\citenamefont {K{\"o}rkel}(2002)}]{Koerkel2002}%
  \BibitemOpen
  \bibfield  {author} {\bibinfo {author} {\bibfnamefont {S.}~\bibnamefont
  {K{\"o}rkel}},\ }\emph {\bibinfo {title} {Numerische {M}ethoden f\"ur
  Optimale {V}ersuchsplanungsprobleme bei nichtlinearen {DAE}-{M}odellen}},\
  \href {http://www.koerkel.de} {Ph.D. thesis},\ \bibinfo  {school}
  {Universit\"at {H}eidelberg}, \bibinfo {address} {Heidelberg} (\bibinfo
  {year} {2002})\BibitemShut {NoStop}%
\bibitem [{\citenamefont {Gill}, \citenamefont {Murray},\ and\ \citenamefont
  {Saunders}()}]{Gill2002}%
  \BibitemOpen
  \bibfield  {author} {\bibinfo {author} {\bibfnamefont {P.}~\bibnamefont
  {Gill}}, \bibinfo {author} {\bibfnamefont {W.}~\bibnamefont {Murray}}, \ and\
  \bibinfo {author} {\bibfnamefont {M.}~\bibnamefont {Saunders}},\ }\bibfield
  {title} {\enquote {\bibinfo {title} {Snopt: An {SQP} algorithm for
  large-scale constrained optimization.}}\ }\href@noop {} {\ }\BibitemShut
  {NoStop}%
\bibitem [{\citenamefont {Bischof}, \citenamefont {Khademi},\ and\
  \citenamefont {Mauer}(1994)}]{Bischof1994}%
  \BibitemOpen
  \bibfield  {author} {\bibinfo {author} {\bibfnamefont {C.}~\bibnamefont
  {Bischof}}, \bibinfo {author} {\bibfnamefont {P.}~\bibnamefont {Khademi}}, \
  and\ \bibinfo {author} {\bibfnamefont {A.}~\bibnamefont {Mauer}},\
  }\href@noop {} {\enquote {\bibinfo {title} {The {ADIFOR} 2.0 system for the
  automatic differentiation of {F}ortran 77 programs},}\ }\bibinfo {type}
  {Technical Report}\ \bibinfo {number} {CRPC--TR94491}\ (\bibinfo
  {institution} {Center for Research on Parallel Computation,Rice University},\
  \bibinfo {address} {Houston, TX},\ \bibinfo {year} {1994})\BibitemShut
  {NoStop}%
\end{thebibliography}
\end{document}